\documentstyle[aps,prl,multicol,amssymb,epsf,psfig]{revtex}
\tightenlines
\title{Entanglement Between Bose-Einstein Condensates
}
\author{Yu Shi\cite{shi}
}
\address{Cavendish Laboratory, University of Cambridge,
 Cambridge CB3 0HE, United Kingdom}
\begin{document}
\draft
\maketitle

\begin{abstract}

For  a Bose condensate in a double-well potential or with two 
Josephson-coupled internal states, the  condensate wavefunction  
is a superposition. Here we consider coupling  two such  Bose 
condensates, and  suggest  the  existence of a joint condensate 
wavefunction, which is in general a superposition of all products 
of the bases condensate wavefunctions of the two condensates. 
The corresponding many-body state is a product of such  superposed 
wavefunctions,  with appropriate symmetrization. These states may  
be potentially useful for  quantum computation. There may be 
robustness and stability due to macroscopic occupation of  a same 
single particle state. The nonlinearity of the condensate wavefunction 
due to  particle-particle interaction may  be utilized to realize 
nonlinear quantum computation, which was  suggested to be capable 
of  solving  NP-complete problems.
  
\end{abstract}

\pacs{PACS numbers: 03.75.Fi, 05.30.-d, 03.67.Lx, 03.65.Ud}

\section*{1. Introduction}

The Hilbert space of the quantum state of 
a composite system  of 
coupled particles  is the tensor product of the Hilbert  spaces  of these 
particles. That is to say, a quantum state of a  composite system is
in general  a superposition of all the products of the bases
 states of the  constituent 
particles. This leads to the situation that 
the   state  of a composite system may not be factorized
to be a direct product of the states of  the constituent 
particles.  This so-called quantum
entanglement   has been a central  attention 
in foundations of quantum mechanics ever since the early 
days~\cite{epr,schrodinger,bell,shi2}. 
Moreover, in recent years, it was found that
superposition and entanglement  of  quantum  states give rise to 
powerful quantum information processing. For example,
a quantum computer is much more efficient than a classical computer in 
solving certain problems such as factoring large numbers~\cite{shor,bennett}.
It is, however, 
highly challenging to physically build a quantum computer, a major 
obstacle being decoherence, i.e. 
the  fragility of a superposed or entangled  state towards coupling with the
environment, although  the quantum error-correction code  diminishes this
difficulty~\cite{steane}. Stimulated by both its  power and the
challenge in its physical implementations, the field of quantum information is 
attracting more and more  attention. 

Here we extend the   consideration to coupled many-particle 
systems, each of which is composed of identical particles,  
but the coupled systems are  distinguishable with each other. Specifically,
we consider coupled 
Bose condensates, each of which is in a double-well potential 
or has two Josephson-coupled internal states (note that 
we refer to   Josephson-coupled condensates as
{\em one} condensate, since in this case,
 all the  particles, even though from different
sources originally, are overlapping identical particles~\cite{kiefer}).
Because of coupling between each particle in one condensate and each
particle in the other condensate, we consider the mean-field  
many-body state which is a product of
copies of a same wavefunction  of  inter-condensate pair of particles,  with 
symmetrization.  
 With the  flexibility of manipulation,
this situation may  be  realizable in the context of 
Bose-Einstein condensation  of trapped atoms~\cite{an,walls,dalfovo,huang1}.
 
A Bose condensate is well described by the condensate wavefunction, which, 
in a mean field theory, is the single particle wavefunction 
in which the  condensation occurs. It can also be  defined as the spontaneous 
gauge symmetry breaking (SGSB)
of the field operator, or  through  the off-diagonal
long-range order (ODLRO)
 of the one-particle reduced density matrix.  
It is a superposition for a condensate in a double-well potential or with
two Josephson-coupled internal states. 
In our consideration of the
coupled condensates, the  wavefunction of each inter-condensate pair of
particles becomes the  joint  condensate 
wavefunction, which can  entangle the condensate  wavefunctions of
individual condensates. 

With superposition and entanglement, it may be possible to use
condensate wavefunction to 
implement a qubit in quantum computation. 
The first possible  advantage is the  intrinsic  robustness and stability.
It may  be viewed as 
a natural realization of a  fault tolerance prescription
based on  symmetrization~\cite{barenco}. To have the best  accuracy,
each Bose system had better be non-interacting. In principle,
this is possible  since the interaction can  be tuned in the case
of  atomic condensate~\cite{castin}.  On the other hand,
it was noted that  incorporating nonlinear 
evolution in   quantum computation leads  to computational power of
solving  NP-complete and \#P problems in polynomial time~\cite{abram}.
This result has remained as an academic curiosity, since
quantum mechanics is linear. Now,
for a weakly interacting  Bose system,  
the condensate wavefunction
satisfies a nonlinear Schr\"{o}dinger equation, often called 
Gross-Pitaevskii equation~\cite{gross}.  Thus  Bose  condensates  
may  make it possible to implement   nonlinear quantum 
computation, for the sake of special power. 
Therefore Bose statistics  may be  
a resource of both   fault  tolerance and  computational power.
We will illustrate the ideas in terms of Bose-Einstein 
condensation of trapped atoms, mainly 
using  a condensate in a double-well potential, with 
the two bits  represented by the  condensate wavefunctions
localized at the two wells. We also discuss
a condensate with
two Josephson-coupled internal states, which might  encode
the bits.  But there
are difficulties due to coupling with the motional degree of freedom 
and the nonlinearity.  However,  the physical problems
are still  interesting.

The paper is organized in the following way. Section 2 is an overview
of the concept of  condensate wavefunction. In Appendix A, we 
give   a   derivation of Gross-Pitaevskii equation from 
ODLRO.  Bose-Einstein   condensation in a double-well
potential is a prototype in discussions from Secs. 3 to 5. 
 As a prelude, superposition of condensate
wavefunction is discussed in Subsec. 3.1.  
In Subsec. 3.2, we  consider   coupling  
two different condensates, and 
discuss possible entanglement between them. 
Many-body
Hamiltonian and the  equation of motion of the joint condensate wavefunction
 are given in Sec. 4. 
Section 5  contains illustrative schemes
of one-bit and two-bit operations  
in terms of condensates in  double-well potentials. 
In Sec. 6, we discuss, in parallel 
to Secs. 3 to 5, the case of spinor condensate,  which has two
components with different internal states. We suggest that coupling two
spinor condensates gives rise to a four-component condensate wavefunction. 
Section 7 contains some additional remarks.  A  summary is given in 
Sec. 8.

\section*{2. Introducing  Condensate Wavefunction}

In the following,  we review  three approaches to the condensate wavefunction. 

(i) In the mean field theory, which 
becomes exact in the  absence of the interaction, 
the Bose-condensed state is  in a Hatree form, 
\begin{equation}
\Psi(\bbox{r_1},\cdots,\bbox{r_N})
=
\phi(\bbox{r_1})\cdots\phi(\bbox{r_N}), \label{mean}
\end{equation}
where $\phi(\bbox{r_i})$ is the  single particle state, which
turns out to be the condensate
wavefunction. One can define 
\begin{equation}
\Phi(\bbox{r})=\sqrt{N}\phi(\bbox{r}). \label{nn}
\end{equation}

(ii) In the 
 approach of SGSB~\cite{goldstone,anderson,leggett}, 
\begin{equation}
\Phi(\bbox{r})=\langle\hat{\psi}(\bbox{r})\rangle,
\end{equation}
where $\hat{\psi}(\bbox{r})$ is the boson field operator.
The particle number is not conserved.
One  may use  (\ref{nn}) to define $\phi(\bbox{r})$,  with $N$ now 
being mean particle number. 
  
(iii) The   general criterion for Bose-Einstein condensation
is the ODLRO  
of the one-particle reduced density matrix~\cite{penrose,yang},
\begin{equation}
\langle\bbox{r'}|\hat{\rho}_1|\bbox{r}\rangle=
\langle \hat{\psi}(\bbox{r'})\hat{\psi}^{\dagger}(\bbox{r})\rangle 
=Tr[\hat{\rho}\hat{\psi}(\bbox{r'})\hat{\psi}^{\dagger}(\bbox{r})],
\end{equation}
where $\hat{\rho}$
 is the density matrix. ODLRO  can be expressed as 
\begin{equation}
\langle\bbox{r'}|\hat{\rho}_1|\bbox{r}\rangle
\rightarrow 
\Phi(\bbox{r'})\Phi(\bbox{r})^* \neq 0 \label{od}
\end{equation}
as $|\bbox{r}-\bbox{r'}|\rightarrow \infty$. 
This applies to both particle number conservation and non-conservation cases.
In case of  particle number conservation,  
$\Phi(\bbox{r})=\sqrt{\lambda}\Phi_0(\bbox{r})$, where $\Phi_0(\bbox{r})$
is the eigenfunction of $\hat{\rho}_1$ with the
 largest eigenvalue  $\lambda$, of the order $o(N)$.
  In case of particle number non-conservation, 
$\Phi(\bbox{r})= \langle\hat{\psi}(\bbox{r})\rangle$.
When  $N\rightarrow\infty$, there is no practical difference in using
 the particle number  non-conserved 
 state or the particle number conserved   state. 

The many-body  Hamiltonian is 
\begin{equation}
\hat{H} = \sum_{i=1}^{N} \hat{h}(\bbox{r}_i)+ 
\sum_{i<j} U(\bbox{r}_i-\bbox{r}_j), \label{oph}
\end{equation}
where 
\begin{equation}
\hat{h}(\bbox{r})= -\frac{\hbar^2\bigtriangledown_{\bbox{r}}^2}{2m}+
V(\bbox{r}) \label{one}
\end{equation}
is the single particle Hamiltonian,  $U(\bbox{r}_i-\bbox{r}_j)$ 
is the particle-particle interaction,   $m$ is 
 the particle mass, $V(\bbox{r})$ is the external potential, 
e.g.  the trapping potential in  case  of the trapped atoms. 
In terms of the field operator, the many-body Hamiltonian is 
\begin{equation}
{\cal H}=\int d\bbox{r}\hat{\psi}^{\dagger}(\bbox{r})\hat{h}(\bbox{r})
\hat{\psi}(\bbox{r})
+\frac{1}{2}\int \int d\bbox{r}_1d\bbox{r}_2 \hat{\psi}^{\dagger}(\bbox{r}_1)
\hat{\psi}^{\dagger}(\bbox{r}_2)U(\bbox{r}_1-\bbox{r}_2) \hat{\psi}(\bbox{r}_2)
 \hat{\psi}(\bbox{r}_1), \label{h1}
\end{equation}
which  leads to the equation of motion of $\hat{\psi}(\bbox{r})$,
\begin{equation}
i\hbar\frac{\partial\hat{\psi}(\bbox{r},t)}{\partial t}=
[-\frac{\hbar^2\bigtriangledown^2_{\bbox{r}}}{2m}+V(\bbox{r})
+\int \hat{\psi}^{\dagger}(\bbox{r}',t)
U(\bbox{r}-\bbox{r}')\hat{\psi}(\bbox{r}',t)
d\bbox{r}']\hat{\psi}(\bbox{r},t).
 \label{ns}
\end{equation}
With weak interaction, one can use  
 s-wave approximation $U(\bbox{r}-\bbox{r}')=g\delta(\bbox{r}-\bbox{r}')$, 
with $g=4\pi\hbar^2\eta/m$, where $\eta$ is the s-wave 
scattering length. Under the  SGSB ansatz, one  replaces 
$\hat{\psi}(\bbox{r},t)$ with  $\sqrt{N}\phi$ to 
 obtain 
 the Gross-Pitaevskii Equation  for the condensate wavefunction,
\begin{equation}
i\hbar\frac{\partial \phi(\bbox{r},t)}{\partial t}=
(-\frac{\hbar^2\bigtriangledown^2}{2m}+V(\bbox{r})
+gN|\phi(\bbox{r},t)|^2)\phi(\bbox{r},t). \label{gp}
\end{equation}
It was    shown    that 
the ground state energy and density given by  Gross-Pitaevskii equation 
become exact as the particle number $N$ tends to be infinity while 
$N\eta$ is fixed~\cite{lieb}, and  the error 
is about $1\%$ under current experimental condition \cite{tim,dalfovo}. 

Gross-Pitaevskii equation can also be obtained by using the Hatree
wavefunction (\ref{mean}) in  the many-body 
Schr\"{o}dinger equation, with the Hamiltonian (\ref{oph})~\cite{gross}.
It can also be  obtained phenomenologically  from a
Ginzburg-Landau functional~\cite{dalfovo}.
In Appendix A, we give a derivation  from  ODLRO.  

\section*{3. Superposition and Entanglement of Condensate Wavefunctions}

\subsection*{3.1. Superposition of condensate wavefunctions
}

In discussing Josephson-like  effect, e.g. 
a condensate  in a  symmetric double-well  potential, a widely used ansatz is 
that the total condensate wavefunction is a 
superposition of  two bases  wavefunctions, with time-dependent 
coefficients~\cite{milburn,smerzi}.
As a prelude to the next subsection, here 
we make some discussions based on 
the construction of the
field operator and then imposing SGSB or ODLRO.

The basis set of the 
single particle wavefunctions is $\{\phi_{\alpha,n}\}$, where $\alpha$ 
denotes the 
energy level, $n=0,1$ denotes the two wells. Because of finiteness of
the barrier, 
$|\int \phi_{\alpha,0}^*(\bbox{r})\phi_{\alpha,1}(\bbox{r})d\bbox{r}|^2
=\epsilon  << 1$, i.e. the two states are nearly orthogonal though not
exactly so. 
The field operator can be constructed as
\begin{equation}
\hat{\psi}(\bbox{r},t)=\sum_{n=0,1} \sum_{\alpha}
\phi_{\alpha,n}(\bbox{r})\hat{a}_{\alpha,n}(t),
\end{equation}
where $\hat{a}_{\alpha,n}(t)$
is the annihilation operator corresponding to 
the single particle state $\phi_{\alpha,n}(\bbox{r})$. Therefore
\begin{equation}
\hat{\psi}(\bbox{r},t)=  \hat{\psi}_0(\bbox{r},t)+\hat{\psi}_1(\bbox{r},t),
  \label{fs}
\end{equation}
where  
\begin{equation}
\hat{\psi}_n(\bbox{r},t)=
\sum_{\alpha}\phi_{\alpha,n}(\bbox{r})\hat{a}_{\alpha,n}(t), 
\end{equation}
with $n=0,1$.
By making SGSB average of  (\ref{fs}), we have
\begin{equation}
\Phi(\bbox{r},t) = \Phi_0(\bbox{r},t)+ \Phi_1(\bbox{r},t), \label{sup}
\end{equation}
with $\Phi(\bbox{r},t)=\langle \hat{\psi}(\bbox{r},t) \rangle$, 
$\Phi_n(\bbox{r},t)= \langle \hat{\psi}_n(\bbox{r},t)\rangle$.  
Therefore,  a general condensate wavefunction is a
superposition of  the condensate wavefunctions corresponding to the
two wells~\cite{zapata}. 

A  justification   can also be made   in terms of ODLRO.
With (\ref{fs}), the one-particle reduced density matrix is 
\begin{equation}
\langle \bbox{r}'|\hat{\rho}_1| \bbox{r}\rangle =  
\langle\bbox{r}',0|\hat{\rho}_1|\bbox{r},0\rangle
+\langle\bbox{r}',0|\hat{\rho}_1|\bbox{r},1\rangle
+\langle\bbox{r}',1|\hat{\rho}_1|\bbox{r},0\rangle
+\langle\bbox{r}',1|\hat{\rho}_1|\bbox{r},1\rangle,
\end{equation}
where 
$\langle\bbox{r}',n'|\hat{\rho}_1|\bbox{r},n\rangle
= \langle\hat{\psi}_{n'}(\bbox{r}',t) 
\hat{\psi}^{\dagger}_{n}(\bbox{r},t)\rangle$,
with $n=0,1$, $n'=0,1$.
The existence of ODLRO implies
%
$$\langle \bbox{r}',n'|\rho_1|\bbox{r},n\rangle
\rightarrow  \Phi_{n'}(\bbox{r}',t)\Phi_n^*(\bbox{r},t).$$
%
Hence
%
$$\langle \bbox{r}'| \hat{\rho}_1| \bbox{r}\rangle \rightarrow 
(\Phi_0(\bbox{r}',t)+\Phi_1(\bbox{r}',t))
(\Phi_0(\bbox{r},t)+\Phi_1(\bbox{r},t))^*,  $$
%
which indicates the existence of 
the superposed condensate wavefunction 
$\Phi(\bbox{r},t)=\Phi_0(\bbox{r},t)+\Phi_1(\bbox{r},t)$.

In a mean field theory, each particle occupies the single particle ground
state, hence the condensate wavefunction is given by the 
superposition of the two single particle ground states at the two wells. 
This is called  two-mode approximation in \cite{milburn}. So 
\begin{equation}
\phi(\bbox{r},t) \equiv \frac{\Phi(\bbox{r},t)}{\sqrt{N}}
=c_0(t)\phi_{\alpha_0,0}(\bbox{r})+
c_1(t)\phi_{\alpha_0,1}(\bbox{r}), \label{pp}
\end{equation}
where $\phi_{\alpha_0,n}(\bbox{r})$ is the single particle ground state
at well $n$.  
The many-body ground state is~\cite{leggett}  
\begin{equation}
\Psi (\bbox{r}_1,\cdots,\bbox{r}_N,t) 
=\phi(\bbox{r}_1,t)\cdots 
\phi(\bbox{r}_N,t). 
\label{ms2}
\end{equation}
For brevity, we write (\ref{pp}) as
\begin{equation}
\phi(\bbox{r},t)=c_0(t)\phi_{0}(\bbox{r})+c_1(t)\phi_{1}(\bbox{r}).
 \label{bre}
\end{equation}

\subsection*{3.2. Entanglement of condensate wavefunctions }

Consider   two coupled Bose systems  $a$ and $b$,  say,   in  a  double-well 
potential (Fig. 1). For simplicity but without lose of generality, suppose 
each consists of $N$ identical  particles.
 For system $a$, the wells are denoted as $n^a$, while for
system $b$, the wells are denoted as $n^b$.
These two Bose systems should  be non-overlapping if they are 
composed of a same kind of  particles, 
but can be mixed if they are composed of two different kinds of particles.

The existence of a joint order parameter can be seen in general by
considering the product of the two field operators. 
\begin{equation}
\hat{\psi}(\bbox{r}^a,t)\hat{\psi}(\bbox{r}^b,t)=
\sum_{n^a}\sum_{n^b} \hat{\psi}_{n^a}(\bbox{r}^a,t)
\hat{\psi}_{n^b}(\bbox{r}^b,t), 
\end{equation}
Imposing SGSB, one has 
\begin{equation}
\Phi(\bbox{r}^a,\bbox{r}^b,t)
=\sum_{n^a}\sum_{n^b}\Phi_{n^a,n^b}(\bbox{r^a},\bbox{r^b},t),
\label{psi1}
\end{equation}
where 
$\Phi(\bbox{r}^a,\bbox{r}^b,t)\equiv\langle \hat{\psi}(\bbox{r}^a,t)
\hat{\psi}(\bbox{r}^b,t)\rangle$, 
$\Phi_{n^a,n^b}(\bbox{r}^a,\bbox{r}^b,t)\equiv
 \langle \hat{\psi}_{n^a}(\bbox{r}^a,t) \hat{\psi}_{n^b}(\bbox{r}^b,t)\rangle$.
Because of coupling between $a$ and $b$, 
 in general the many-body state cannot be factorized to be a direct product of
 that of $a$ and that of $b$, and there is interaction energy between
$a$ and $b$. Consequently, 
 $\Phi(\bbox{r}^a,\bbox{r}^b,t)$ cannot be factorized
to be a product of a condensate wavefunction of system $a$ and 
a condensate wavefunction of system $b$. 

The justification can also be made in terms of ODLRO. 
One can define a  one-particle-pair reduced density matrix, 
\begin{eqnarray}
\langle {\bbox{r}^a}',{\bbox{r}^b}'|\hat{\rho}_1|\bbox{r}^a,\bbox{r}^b\rangle 
&\equiv&  \langle \hat{\psi}({\bbox{r}^a}',t)\hat{\psi}({\bbox{r}^b}',t)
\hat{\psi}^{\dagger}(\bbox{r}^b,t)
\hat{\psi}^{\dagger}(\bbox{r}^a,t)\rangle \nonumber \\
&= &\sum_{{n^a}'}\sum_{{n^b}'}\sum_{n^a}\sum_{n^b}
\langle {\bbox{r}^a}',{n^a}',{\bbox{r}^b}',{n^b}'|\hat{\rho}_1|
\bbox{r}^a,n^a,\bbox{r}^b,n^b \rangle, \label{o1}
\end{eqnarray}
where 
$\langle {\bbox{r}^a}',{n^a}',{\bbox{r}^b}',{n^b}'|\hat{\rho}_1|
\bbox{r}^a,n^a,\bbox{r}^b,n^b \rangle
\equiv \langle \hat{\psi}_{{n^a}'}({\bbox{r}^a}',t)
 \hat{\psi}_{{n^b}'}({\bbox{r}^b}',t)
\hat{\psi}_{n^b}^{\dagger}(\bbox{r}^b,t)
\hat{\psi}_{n^a}^{\dagger}(\bbox{r}^a,t)\rangle$.
With  ODLRO,
\begin{equation}
\langle \hat{\psi}_{{n^a}'}({\bbox{r}^a}',t)
 \hat{\psi}_{{n^b}'}({\bbox{r}^b}',t)
\hat{\psi}_{n^b}^{\dagger}(\bbox{r}^b,t)
\hat{\psi}_{n^a}^{\dagger}(\bbox{r}^a,t)\rangle
\rightarrow
\Phi_{{n^a}',{n^b}'}({\bbox{r}^a}',{\bbox{r}^b}',t)
\Phi_{n^a,n^b}^{\dagger}(\bbox{r}^a,\bbox{r}^b,t).
\end{equation}
Consequently,  (\ref{o1}) approaches
$$[\sum_{{n^a}'}\sum_{{n^b}'}
\Phi_{{n^a}',{n^b}'}({\bbox{r}^a}',{\bbox{r}^b}',t)]
[\sum_{n^a}\sum_{n^b}
\Phi_{n^a,n^b}(\bbox{r}^a,\bbox{r}^b,t)]^*,
$$
implying that there exists a joint   condensate wavefunction
$\sum_{n^a,n^b}\Phi_{n^a,n^b}(\bbox{r}^a,\bbox{r}^b,t)$.

For convenience, we may write (\ref{psi1}) as
\begin{equation}
\phi(\bbox{r}^a,\bbox{r}^b,t)
=\sum_{n^a}\sum_{n^b}
\phi_{n^a,n^b}(\bbox{r}^a,\bbox{r}^b,t),
\label{psi3}
\end{equation}
with $\phi(\bbox{r}^a,\bbox{r}^b,t)=\Phi(\bbox{r}^a,\bbox{r}^b,t)/N$,
$\phi_{n^a,n^b}(\bbox{r^a},\bbox{r^b},t)=
\Phi_{n^a,n^b}(\bbox{r^a},\bbox{r^b},t)/N$.

Consider the following
Bose-condensed state, 
\begin{equation}
\Psi(\bbox{r}^{a}_{1},\cdots,\bbox{r}^{a}_N;
\bbox{r}^{b}_{1},\cdots,\bbox{r}^{b}_{N};t) =\frac{1}{\sqrt{N!}}
\sum_{P}
[\phi_{\alpha^a_0,\alpha^b_0}(\bbox{r}^{a}_{1},\bbox{r}^{b}_{P1},t) \cdots 
\phi_{\alpha^a_0,\alpha^b_0}(\bbox{r}^{a}_{N},\bbox{r}^{b}_{PN},t)], 
\label{en}
\end{equation}
where 
\begin{equation}
\phi_{\alpha^a_0,\alpha^b_0}(\bbox{r}^{a},\bbox{r}^{b},t)
=\sum_{n^a}\sum_{n^b}
C_{n^a n^b}(t) \phi_{\alpha^a_0,n^a}(\bbox{r}^a)\phi_{\alpha^b_0,n^b}
(\bbox{r}^{b}).
\label{psi}
\end{equation}
The symmetrization in  state (\ref{en}) represents all possible ways
of pairing. 
The systems  Bose-condense into a coupled single particle 
pair state, and (\ref{psi}) can be used as  the joint condensate wavefunction. 
In this state,  particles in a same system occupy a same ground state, while
maintain  coupling with particles in the other system.  It is a natural
generalization of the ansatz (\ref{ms2}). 
However, it remains to be rigorously examined whether this state is indeed 
the  many-body ground state.

Like using  two-mode approximation for a single condensate, 
one may make  two-mode approximation for each condensate, and consider
the coupling between them.   Then,
in consistent with the above mean-field many-body state, 
the joint condensate wavefunction   is  
\begin{equation}
\phi(\bbox{r}^{a},\bbox{r}^{b},t)\equiv
\frac{ \langle \hat{\psi}(\bbox{r}^a,t)
\hat{\psi}(\bbox{r}^b,t)\rangle}{N^2}
 =\phi_{\alpha^a_0,\alpha^b_0}(\bbox{r}^{a},\bbox{r}^{b},t) 
\nonumber
\end{equation}
For brevity, we write 
\begin{equation}
\phi(\bbox{r}^{a},\bbox{r}^{b},t)
=\sum_{n^a}\sum_{n^b}
C_{n^a n^b}(t) \phi_{n^a}(\bbox{r}^a)\phi_{n^b}(\bbox{r}^{b}). 
\label{add}
\end{equation}

\section*{4. Many-body Hamiltonian and Equation of Motion }

For  two coupled Bose systems $a$ and $b$,
the general form of the Hamiltonian  is 
\begin{equation}
\hat{H} = \sum_{i=1}^{N} \hat{h}(\bbox{r}_i^a)+ 
\sum_{i<j} U(\bbox{r}_i^a-\bbox{r}_j^a)
+\sum_{k=1}^{N} \hat{h}(\bbox{r}_k^b)+ 
\sum_{k<l} U(\bbox{r}_k^b-\bbox{r}_l^b)
+\sum_{i,k} W(\bbox{r}_i^a-\bbox{r}_k^b),  \label{many}
\end{equation}
where  $\hat{h}(\bbox{r}_i^a)$ and $\hat{h}(\bbox{r}_k^b)$ are  
single particle Hamiltonians of a particle in $a$ and $b$, respectively,
as given in (\ref{one}).
 $U(\bbox{r}_i^a-\bbox{r}_j^a)$ 
is the particle-particle interaction within $a$, while 
$U(\bbox{r}_k^b-\bbox{r}_l^b)$ is the interaction  within $b$.
 $W(\bbox{r}_i^a-\bbox{r}_k^b)$ is the interaction between a particle in
$a$ and a particle in $b$. 
The field theoretic   Hamiltonian can be  written as  
\begin{equation}
{\cal H}={\cal H}_1+{\cal H}_2+{\cal H}_3+{\cal H}_4+{\cal H}_5,
\label{mn}
\end{equation}
with
\begin{equation}
{\cal H}_1=\int d\bbox{r}^a
\hat{\psi}^{\dagger}(\bbox{r}^a)\hat{h}(\bbox{r}^a)
\hat{\psi}(\bbox{r}^a), \label{h0}
\end{equation}
\begin{equation}
{\cal H}_2=\int d\bbox{r}^b
\hat{\psi}^{\dagger}(\bbox{r}^b)\hat{h}(\bbox{r}^b)
\hat{\psi}(\bbox{r}^b),\label{h2}
\end{equation}
\begin{equation}
{\cal H}_3
=\frac{1}{2}\int\int  d\bbox{r}^a_1   
d\bbox{r}^a_2 
\hat{\psi}^{\dagger}(\bbox{r}^a_1)
\hat{\psi}^{\dagger}(\bbox{r}^a_2)
U(\bbox{r}_1^a-\bbox{r}_2^a)
\hat{\psi}(\bbox{r}^a_2)
\hat{\psi}(\bbox{r}^a_1) \label{h3}
\end{equation}
\begin{equation}
{\cal H}_4
=\frac{1}{2}\int\int  d\bbox{r}^b_1 d\bbox{r}^b_2
\hat{\psi}^{\dagger}(\bbox{r}^b_1)
\hat{\psi}^{\dagger}(\bbox{r}^b_2)
U(\bbox{r}_1^b-\bbox{r}_2^b)
\hat{\psi}(\bbox{r}^b_2)
\hat{\psi}(\bbox{r}^b_1) \label{h4}
\end{equation}
\begin{equation}
{\cal H}_5
=
\int\int  d\bbox{r}^a d\bbox{r}^b  
\hat{\psi}^{\dagger}(\bbox{r}^a)
\hat{\psi}^{\dagger}(\bbox{r}^b)
W(\bbox{r}^a-\bbox{r}^b)
\hat{\psi}(\bbox{r}^b)
\hat{\psi}(\bbox{r}^a),  \label{h5}
\end{equation}

From the   Hamiltonian (\ref{mn}), 
  the equation of motion of 
 $\hat{\psi}(\bbox{r}^a)\hat{\psi}(\bbox{r}^b)$ is obtained as 
\begin{equation}
\begin{array}{rl}
i\hbar\frac{\partial [\hat{\psi}(\bbox{r}^a,t)\hat{\psi}(\bbox{r}^b,t)]}
{\partial t}=&
[\hat{h}(\bbox{r}^a)+\hat{h}(\bbox{r}^b)+
\int d{\bbox{r}^a}'
\hat{\psi}^{\dagger}({\bbox{r}^a}',t)U(\bbox{r}^a-{\bbox{r}^a}')
\hat{\psi}({\bbox{r}^a}',t) \\
&+\int d{\bbox{r}^b}'
\hat{\psi}^{\dagger}({\bbox{r}^b}',t)U(\bbox{r}^b-{\bbox{r}^b}')
\hat{\psi}({\bbox{r}^b}',t)  \\
&+\int d{\bbox{r}^a}'
\hat{\psi}^{\dagger}({\bbox{r}^a}',t)W({\bbox{r}^a}'-\bbox{r}^b)
\hat{\psi}({\bbox{r}^a}',t) \\
&+\int d{\bbox{r}^b}'
\hat{\psi}^{\dagger}({\bbox{r}^b}',t)W(\bbox{r}^a-{\bbox{r}^b}')
\hat{\psi}({\bbox{r}^b}',t)]\hat{\psi}(\bbox{r}^a,t)\hat{\psi}(\bbox{r}^b,t),
\end{array}
  \label{eom3}
\end{equation}
from which,  using
$\int d\bbox{r}^a 
\hat{\psi}^{\dagger}(\bbox{r}^a,t)\hat{\psi}(\bbox{r}^a,t)
=\int d\bbox{r}^b
\hat{\psi}^{\dagger}(\bbox{r}^b,t)\hat{\psi}(\bbox{r}^b,t)=N$,
 one obtains the equation of motion for the 
joint  condensate wavefunction  
$\phi(\bbox{r}^a,\bbox{r}^b,t)$, as
a generalization  of the  Gross-Pitaevskii equation, 
\begin{equation}
\begin{array}{rl}
i\hbar\frac{\partial \phi(\bbox{r}^a,\bbox{r}^b,t)}{\partial t} =&
[\hat{h}(\bbox{r}^a)+\hat{h}(\bbox{r}^b)
+N\int\int d{\bbox{r}^a}'d{\bbox{r}^b}'
|\phi ({\bbox{r}^a}',{\bbox{r}^b}',t)|^2 U(\bbox{r}^a-{\bbox{r}^a}')
 \\
&+N\int\int d{\bbox{r}^a}'d{\bbox{r}^b}'
|\phi ({\bbox{r}^a}',{\bbox{r}^b}',t)|^2 U(\bbox{r}^b-{\bbox{r}^b}')
\\
&
+N
\int\int d{\bbox{r}^a}'d{\bbox{r}^b}'|\phi ({\bbox{r}^a}',{\bbox{r}^b}',t)|^2 
W({\bbox{r}^a}'-\bbox{r}^b)  \\
&
+N
\int\int d{\bbox{r}^a}'d{\bbox{r}^b}'|\phi ({\bbox{r}^a}',{\bbox{r}^b}',t)|^2 
W(\bbox{r}^a-{\bbox{r}^b}')] \phi(\bbox{r}^a,\bbox{r}^b,t).
\end{array} \label{couple}
\end{equation}

This equation may  also be obtained 
directly from the many-body Schr\"{o}dinger equation, using (\ref{en}) and
(\ref{many}). 

\section*{5. Quantum computation with double-well  condensates}

\subsection*{5.1. Possible robustness} 

If a qubit is implemented as  a Bose condensate, then 
there is possible robustness and stability  due to   the macroscopic 
occupation of a same single particle state.
For example,
consider an excited state where only one particle is away from the
ground state, for distinguishable particles there are $N$ possibilities, 
while there is only one possibility when all the particles are 
{\em identical}~\cite{kerson}.  For a single condensate, the Bose-condensed
 state
is a product of the same  single particle state. For two  coupled  condensates,
the Bose-condensed state is now suggested to be 
a product of the same single particle pair state,
with symmetrization over all possible ways of pairing. 
In general, symmetrization always  needs to be made on the many-body state. 
This  reduces  the error probability.  
Symmetrization has been studied as a way of reducing 
error in 
quantum computation, which was found to suppress the error probability 
by $1/N$~\cite{barenco}. Bose condensation can be viewed as
  a natural realization of this prescription. 
The error reduction 
due to symmetrization may help one to understand why the single particle state
emerges out as a   macroscopic wavefunction. 
 The robustness of condensate wavefunction
was demonstrated in the interference experiments~\cite{phase}.
Nevertheless, the condensate wavefunction brings  the 
issue of   phase diffusion,  which may cause error.  

\subsection*{5.2. NP-complete problems and Nonlinear Quantum Computation}

The class of NP-complete problems is a foundation of the computational 
complexity theory. It  includes thousands of 
practically interesting problems, such as travelling salesman,
satisfiability, etc. NP stands for ``non-deterministic polynomial time''. 
NP-complete problems are those for which a potential solution can be 
verified in polynomial time, yet finding a solution appears to require
exponential time in the worst case.
The completeness means that if  an efficient, i.e. polynomial-time, 
algorithm could be found for solving one of these problems, 
one would immediately have an efficient algorithm for all
NP-complete problems.  A fundamental conjecture in classical computation 
is that no such an efficient algorithm exists. 
Abrams and Lloyd 
  found that  with   nonlinearity, a quantum computer can solve
NP-complete problems by  efficiently 
determining  if  there exists an  $x$ for which $f(x)=1$, and 
 can solve \#P problems  by efficiently determining the number of 
solutions \cite{abram}. Their  algorithm  is based on  one or two
 one-bit nonlinear gates, 
  together with linear gates. However,  
it is an experimental fact that fundamental  quantum mechanics
is linear to the available accuracy, while
nonlinear fundamental quantum  theory \cite{weinberg} usually 
violates the second law of thermodynamics \cite{peres} and 
the theory of  relativity  \cite{gisin}. 

Now, because the condensate wavefunction is nonlinear when there is interaction
between  particles, while it has been shown that the  condensate wavefunctions
of different Bose condensates can be entangled, 
we propose  that condensate qubit may   be used to realize
 the  nonlinear quantum computing, and thus deal with  NP-complete 
and \#P problems.  It might be also  possible to use it to 
go around some   constraints in quantum information processing   originated
in  linearity of quantum mechanics.
For a Bose condensate of trapped atoms, 
the atom-atom interaction, and thus
the nonlinearity of condensate wavefunction,  
can be tuned. Therefore in principle
one  may construct  linear  gates by turning off the nonlinearity,
and nonlinear gates by turning on the nonlinearity. 
 The nonlinearity of  Gross-Pitaevskii equation   is just of
Weinberg-type   \cite{weinberg}, which is  used in the
 algorithms in \cite{abram}.

As explained in Sec.~2, in general
the condensate wavefunction is not the total 
 pure state  of a closed system, hence of course 
 its nonlinearity has nothing to do with,
and does not share the problems of, nonlinear quantum mechanics. 
In the absence of the interaction, the condensate wavefunction 
reduces to the pure state of a single particle, and, consistently,
  the nonlinearity disappears. 

\subsection*{5.3. One-bit operation}

We  illustrate  how to implement the one-bit gates,
in terms of a Bose  condensate of trapped atoms 
in   a  symmetric  double-well  trapping  potential $V(\bbox{r})$ (Fig. 1).
This is essentially  the Josephson-like 
effect investigated previously~\cite{milburn}.
We may represent bits  $|0\rangle$ and $|1\rangle$
as the    condensate wavefunctions at the 
two  wells, respectively. Thus  
\begin{equation}
|n\rangle = \int \phi_{n}(\bbox{r})|\bbox{r}\rangle d\bbox{r},
\end{equation}
where $n=0,1$,
$\phi_{n}(\bbox{r})=u(\bbox{r}-\bbox{r}_n)$ is the condensate wavefunction
corresponding to    the  local potential 
$\tilde{V}(\bbox{r}-\bbox{r}_n)$,
which may be  parabolic, 
at the vicinity of the   bottom $\bbox{r}_n$. 
In accordance with Eq.~(\ref{bre}), a  qubit $|q(t)\rangle$ 
is in general  a superposition of $|0\rangle$ and $|1\rangle$,
\begin{equation}
|q(t)\rangle=c_0(t)|0\rangle+c_1(t)|1\rangle$$\doteq$$\left(
 \begin{array}{c}
c_0(t)\\
c_1(t)\end{array}\right),
\end{equation}
where  $\doteq$  denotes the matrix representation, with
$$|0\rangle \doteq \left(
 \begin{array}{c}
1\\
0\end{array}\right),
|1\rangle\doteq\left(
 \begin{array}{c}
0\\
1\end{array}\right).$$

 Gross-Pitaevskii equation leads to
\begin{equation}
i\hbar \frac{\partial}{\partial t}|q(t)\rangle
= \hat{H}|q(t)\rangle,  \label{qun}
\end{equation}
with
\begin{equation}
\hat{H}= E\hat{I}+\Omega\left( \begin{array}{cc}
0 & 1\\
1 & 0 \end{array} \right)
+\kappa N \left( \begin{array}{cc}
|c_0|^2&0\\
0&|c_1|^2 \end{array} \right), \label{bit}
\end{equation}
where 
\begin{equation}
\Omega=\int d\bbox{r}\phi_0^*(\bbox{r})
[V(\bbox{r})-\tilde{V}(\bbox{r}-\bbox{r_0})]
\phi_1(\bbox{r})
\end{equation}
 represents the Josephson-like  tunnelling effect, 
\begin{equation}
\kappa=g\int d\bbox{r}|\phi_0(\bbox{r})|^4,
\end{equation}
 $E$ is the 
 energy of the basis $|0\rangle$ and $|1\rangle$.   
Therefore, when the s-wave
interaction is turned off, $\kappa=0$, we may
have  an arbitrary one-bit linear  transformation, depending on the
time span $\tau$:
\begin{equation}
|q(\tau)\rangle\rightarrow \exp[-\frac{i}{\hbar}
(E\hat{I}+\Omega\hat{\sigma}_x)\tau]|q(0)\rangle.
\end{equation}
Thus one may  construct  one-bit linear gates.
When  the s-wave  interaction is turned on,   $\kappa\neq 0$,
there is  a twisting rotation in the  state space spanned 
by $|0\rangle$ and $|1\rangle$. By choosing appropriate time span, 
this may  be used to construct  one-bit nonlinear  gates.  

We mention that when the wavefunction $\phi_n(\bbox{r})$ is real,
$\phi_0(\bbox{r})\pm \phi_1(\bbox{r}) $  are orthogonal, and might be
used as qubit. 

\subsection*{5.4. Two-bit Operations}

The nonlinear one-bit gates of  Bose-Einstein condensates may 
be integrated with the linear gates of other qubit carriers, so that  
 the algorithms in  \cite{abram} can be implemented, since only one-bit
 nonlinear  gates are needed there. 
A network of condensates is also possible. 
In the following,  
we investigate the  evolution of two coupled condensates based 
 on a direct interaction, as described in Section 5. 
  We consider  trapped atoms in a double-well potential.
A long range interaction, denoted as $W(\bbox{r}-\bbox{r}')$,  such as 
 dipole-dipole interaction, is a possible basis of 
  the  inter-condensate interaction.  
It  is  considerable either for the magnetic moments of 
trapped atoms with high magnetic moments 
\cite{goral}, or for the electric dipoles induced by strong dc fields
\cite{you}. 
It would be ideal if  the kind of interaction between atoms in different
condensates, for the purpose of coupling, is absent  or somehow canceled 
out  between atoms in a same condensate. 
However, we shall discuss the   general case, since it is interesting
no matter whether it is used for quantum computation. 
The discussion is 
formal, without detailed consideration of 
 the suitable physical conditions. 

In the presence of 
 a  long range interaction $W(\bbox{r}-\bbox{r}')$ 
in addition to the s-wave interaction, one should substitute
  $U(\bbox{r}-\bbox{r}')
=g\delta(\bbox{r}-\bbox{r}')+W(\bbox{r}-\bbox{r}')$ in the  
equation of motion.  
One-bit Hamiltonian (\ref{bit}) is then  modified to  
\begin{equation}
\hat{H}= E\hat{I}+\Omega\left( \begin{array}{cc}
0 & 1\\
1 & 0 \end{array} \right)
+N \left( \begin{array}{cc}
|c_0|^2\kappa+|c_0|^2\mu_1+|c_1|^2\mu_2&0\\
0&\kappa|c_1|^2+|c_1|^2\mu_1+|c_0|^2\mu_2 \end{array} \right), 
\end{equation}
where 
$\mu_1=  
\int W(\bbox{r}-\bbox{r}')|\phi_n(\bbox{r})|^2|\phi_n(\bbox{r}')|^2
d\bbox{r}d\bbox{r}'$
is due to  $W$ within a same well, while $\mu_2=  
\int W(\bbox{r}-\bbox{r}')
 |\phi_n(\bbox{r})|^2 |\phi_{\overline{n}}(\bbox{r}')|^2d\bbox{r}d\bbox{r}'$,
with $\overline{n}=1-n$ , is due to  $W$ within different wells. 
$\mu_1 \gg \mu_2$, since $W$ between atoms 
within a same  well is much larger than that between atoms in
 different wells.

A two-bit gate may be constructed by
putting  together two double-wells, each of which confines a condensate.
They are  close to each other in a face-to-face way (Fig. 1), i.e.,
$|0\rangle_a$ is close to $|0\rangle_b$, and 
$|1\rangle_a$ is close to $|1\rangle_b$. Therefore for 
atoms in different condensates, 
$W$ between, say,  an atom in $|0\rangle_a$ and an atom in  $|0\rangle_b$
is much larger than that between an  atom 
in $|0\rangle_a$ and an atom in $|1\rangle_b$. This is the  
origin  of the
conditional dynamics. 

Substituting Eq.~(\ref{add}) 
to the equation of motion of the total condensate wavefunction,
 Eq. (\ref{couple}),
 with $U(\bbox{r}^i-{\bbox{r}^i}')=
g^i\delta (\bbox{r}^i-{\bbox{r}^i}')+W(\bbox{r}^i-{\bbox{r}^i}')$, ($i=a,b$),
  we obtain  
\begin{equation}
i\hbar\frac{\partial C_{n^a,n^b}}{\partial t} = 
\sum_{n_1^a,n_1^b} C_{n^a_1,n^b_1} (I+II+III+IV+V+VI+VII+VIII), \label{two}
\end{equation}
with 
\begin{equation}
\begin{array}{rl}
I&=\int\int  d\bbox{r}^a d\bbox{r}^b \phi_{n^a}^*(\bbox{r}^a)
\phi_{n^b}^*(\bbox{r}^b)\hat{h}(\bbox{r}^a)\phi_{n^a_1}(\bbox{r}^a)
\phi_{n^b_1}(\bbox{r}^b)   \\ 
&= \delta_{n^b_1,n^b}(E^a\delta_{n^a_1,n^a}
+\Omega^a\delta_{n^a_1,\overline{n}^a}),
\end{array}
\end{equation}
\begin{equation}
\begin{array}{rl}
II&=\int\int  d\bbox{r}^a d\bbox{r}^b \phi_{n^a}^*(\bbox{r}^a)
\phi_{n^b}^*(\bbox{r}^b)\hat{h}(\bbox{r}^b)\phi_{n^a_1}(\bbox{r}^a)
\phi_{{n^b}_1}(\bbox{r}^b) \\
&= \delta_{n^a_1,n^a}(E^b\delta_{n^b_1,n^b}
+\Omega^b\delta_{n^b_1,\overline{n}^b}),
\end{array}
\end{equation}
\begin{equation}
\begin{array}{rl}
III&=g^aN
\int\int\int\int   d\bbox{r}^a d\bbox{r}^b d{\bbox{r}^a}'d{\bbox{r}^b}'
 \phi_{n^a}^*(\bbox{r}^a)\phi_{n^b}^*(\bbox{r}^b)\phi_{n^a_1}(\bbox{r}^a)
\phi_{n^b_1}(\bbox{r}^b) |\phi({\bbox{r}^a}',{\bbox{r}^b}')|^2 
\delta (\bbox{r}^a-{\bbox{r}^a}')  \\ 
&=   \delta_{n^a_1,n^a} \delta_{n^b_1,n^b} g^aN
|C_{n^a,n^b}|^2 \int |\phi_{n^a}(\bbox{r}^a)|^4 d\bbox{r}^a 
   \int |\phi_{n^b}(\bbox{r}^b)|^4 d\bbox{r}^b  \\
&=   \delta_{n^a_1,n^a} \delta_{n^b_1,n^b} g^aN |C_{n^a,n^b}|^2
\kappa^a \kappa^b,
\end{array}
\end{equation}
\begin{equation}
\begin{array}{rl}
IV&=N
\int\int\int\int   d\bbox{r}^a d\bbox{r}^b d{\bbox{r}^a}'d{\bbox{r}^b}'
 \phi_{n^a}^*(\bbox{r}^a)\phi_{n^b}^*(\bbox{r}^b)\phi_{n^a_1}(\bbox{r}^a)
\phi_{n^b_1}(\bbox{r}^b) |\phi({\bbox{r}^a}',{\bbox{r}^b}')|^2 
W(\bbox{r}^a-{\bbox{r}^a}')  \\ 
&=    \delta_{n^a_1,n^a} \delta_{n^b_1,n^b} N
\sum_{{n^a}',{n^b}'}|C_{{n^a}',{n^b}'}|^2 \int |\phi_{n^a}(\bbox{r}^a)|^2 
|\phi_{{n^a}'}({\bbox{r}^a}')|^2 W(\bbox{r}^a-{\bbox{r}^a}') d\bbox{r}^a 
d{\bbox{r}^a}'  \\
&=    \delta_{n^a_1,n^a} \delta_{n^b_1,n^b} N 
\sum_{{n^a}'}(|C_{{n^a}',0}|^2+|C_{{n^a}',1}|^2)\mu_{n^a,{n^a}'},
\end{array}
\end{equation}
\begin{equation}
\begin{array}{rl}
V&=g^bN
\int\int\int\int   d\bbox{r}^a d\bbox{r}^b d{\bbox{r}^a}'d{\bbox{r}^b}'
 \phi_{n^a}^*(\bbox{r}^a)\phi_{n^b}^*(\bbox{r}^b)\phi_{n^a_1}(\bbox{r}^a)
\phi_{n^b_1}(\bbox{r}^b) |\phi({\bbox{r}^a}',{\bbox{r}^b}')|^2 
\delta(\bbox{r}^b-{\bbox{r}^b}')  \\ 
&=   \delta_{n^a_1,n^a} \delta_{n^b_1,n^b} g^b N
|C_{n^a,n^b}|^2 \int |\phi_{n^a}(\bbox{r}^a)|^4 d\bbox{r}^a 
   \int |\phi_{n^b}(\bbox{r}^b)|^4 d\bbox{r}^b \\ 
&= \delta_{n^a_1,n^a} \delta_{n^b_1,n^b} g^b N   
|C_{n^a,n^b}|^2 \kappa^a \kappa^b,
\end{array}
\end{equation}
\begin{equation}
\begin{array}{rl}
VI&=N
\int\int\int\int   d\bbox{r}^a d\bbox{r}^b d{\bbox{r}^a}'d{\bbox{r}^b}'
 \phi_{n^a}^*(\bbox{r}^a)\phi_{n^b}^*(\bbox{r}^b)\phi_{n^a_1}(\bbox{r}^a)
\phi_{n^b_1}(\bbox{r}^b) |\phi({\bbox{r}^a}',{\bbox{r}^b}')|^2 
W(\bbox{r}^b-{\bbox{r}^b}')  \\ 
&=    \delta_{n^a_1,n^a} \delta_{n^b_1,n^b} N 
\sum_{{n^a}',{n^b}'}|C_{{n^a}',{n^b}'}|^2 \int |\phi_{n^b}(\bbox{r}^b)|^2 
|\phi_{{n^b}'}({\bbox{r}^b}')|^2 W(\bbox{r}^b-{\bbox{r}^b}') d\bbox{r}^b
d{\bbox{r}^b}' \\ 
&=    \delta_{n^a_1,n^a} \delta_{n^b_1,n^b} N 
\sum_{{n^b}'}(|C_{0,{n^b}'}|^2+|C_{1,{n^b}'}|^2) \mu_{n^b,{n^b}'},
\end{array}
\end{equation}
\begin{equation}
\begin{array}{rl}
VII&=N
\int\int\int\int   d\bbox{r}^a d\bbox{r}^b d{\bbox{r}^a}'d{\bbox{r}^b}'
 \phi_{n^a}^*(\bbox{r}^a)\phi_{n^b}^*(\bbox{r}^b)\phi_{n^a_1}(\bbox{r}^a)
\phi_{n^b_1}(\bbox{r}^b) |\phi({\bbox{r}^a}',{\bbox{r}^b}')|^2 
W({\bbox{r}^a}'-\bbox{r}^b)  \\ 
&=   \delta_{n^a_1,n^a} \delta_{n^b_1,n^b} N 
\sum_{{n^a}',{n^b}'}|C_{{n^a}',{n^b}'}|^2 \int 
|\phi_{{n^a}'}({\bbox{r}^a}')|^2 |\phi_{n^b}(\bbox{r}^b)|^2  
W({\bbox{r}^a}'-\bbox{r}^b)  d\bbox{r}^b d{\bbox{r}^a}'  \\ 
&= \delta_{n^a_1,n^a} \delta_{n^b_1,n^b}  N  
\sum_{{n^a}'}(|C_{{n^a}',0}|^2+ |C_{{n^a}',1}|^2) \nu_{{n^a}',n^b},
\end{array}
\end{equation}
\begin{equation}
\begin{array}{rl}
VIII&=N
\int\int\int\int   d\bbox{r}^a d\bbox{r}^b d{\bbox{r}^a}'d{\bbox{r}^b}'
 \phi_{n^a}^*(\bbox{r}^a)\phi_{n^b}^*(\bbox{r}^b)\phi_{n^a_1}(\bbox{r}^a)
\phi_{n^b_1}(\bbox{r}^b) |\phi({\bbox{r}^a}',{\bbox{r}^b}')|^2 
W({\bbox{r}^b}'-\bbox{r}^a)  \\ 
&=  \delta_{n^a_1,n^a} \delta_{n^b_1,n^b} N  
\sum_{{n^a}',{n^b}'}|C_{{n^a}',{n_b}'}|^2 \int 
|\phi_{{n^b}'}({\bbox{r}^b}')|^2 |\phi_{n^a}(\bbox{r}^a)|^2  
W({\bbox{r}^b}'-\bbox{r}^a)  d\bbox{r}^a d{\bbox{r}^b}' \\ 
&=  \delta_{n^a_1,n^a} \delta_{n^b_1,n^b} N  
\sum_{{n^b}'}(|C_{0,{n_b}'}|^2+|C_{1,{n_b}'}|^2)\nu_{{n^b}',n^a},
\end{array}
\end{equation}
where $E^i$, $\Omega^i$, and $\kappa^i$, ($i=a,b$) have the same meanings, 
respectively, 
as those quantities without the superscripts,
defined above for a single condensate. $\overline{n}^i=1-n^i$.
For simplicity, we may  set $E^a=E^b$, $\Omega^a=\Omega^b$, 
$g^a=g^b$, $\kappa^a=\kappa^b$. 
$\mu_{n^i,{n^i}'}=\int |\phi_{n^i}(\bbox{r}^i)|^2 
|\phi_{{n^i}'}({\bbox{r}^i}')|^2 W(\bbox{r}^i-{\bbox{r}^i}') d\bbox{r}^i
d{\bbox{r}^i}'$ is due to the interaction $W$ within a same condensate
 qubit. 
$\nu_{n^i,{n^j}'}=\int  |\phi_{n^i}(\bbox{r}^i)|^2  
W({\bbox{r}^j}'-\bbox{r}^i)  |\phi_{{n^j}'}({\bbox{r}^j}')|^2
d\bbox{r}^i d{\bbox{r}^j}'$ is due to the interaction $W$ between different
condensate qubits.
With symmetry,   $\mu_{n^i,{n^i}'}=\mu^i_1$ for $n^i={n^i}'$ while 
$\mu_{n^i,{n^i}'}=\mu^i_2$ for  $n^i \neq {n^i}'$.  $\mu^i_1\gg \mu^i_2$.
 Similarly, 
 $\nu_{n^i,{n^j}'}=\nu_1$ for $n^i={n^j}'$  while 
$\nu_{n^i,{n^j}'}=\nu_2$ for  $n^i \neq {n^j}'$.   $\nu_1 \gg \nu_2$. 
Then Eq. (\ref{two}) can be written as  a matrix equation,
\begin{equation}
 i\hbar\frac{\partial }{\partial t} 
\left(  \begin{array}{c}
C_{00}\\
C_{01}\\
C_{10}\\
C_{11} \end{array} \right)  
=
 \left( \begin{array}{cccc}
G_{00}+N F_{00}&\Omega^b&\Omega^a&0\\
\Omega^b&G_{01}+N F_{01}&0&\Omega^a\\
\Omega^a&0&G_{10}+N F_{10}&\Omega^b\\
0&\Omega^a&\Omega^b&G_{11}+N F_{11}
\end{array} \right)  
 \left( \begin{array}{c}
C_{00}\\
C_{01}\\
C_{10}\\
C_{11} \end{array} \right)  ,  \label{mat}
\end{equation}
where 
\begin{equation}
G_{n^an^b}=E^a+E^b+(g^a+g^b)N\kappa^a\kappa^b|C_{n^an^b}|^2,
\end{equation}
and
\begin{eqnarray}
F_{00}=(\mu_1+\nu_1)(2|C_{00}|^2+|C_{01}|^2+|C_{10}|^2)
+(\mu_2+\nu_2)(|C_{01}|^2+|C_{10}|^2+2|C_{11}|^2),\\
F_{01}=(\mu_1+\nu_2)(|C_{00}|^2+2|C_{01}|^2+|C_{11}|^2)
+(\mu_2+\nu_1)(|C_{00}|^2+2|C_{10}|^2+2|C_{11}|^2),\\
F_{01}=(\mu_1+\nu_2)(|C_{00}|^2+2|C_{10}|^2+|C_{11}|^2)
+(\mu_2+\nu_1)(|C_{00}|^2+2|C_{01}|^2+2|C_{11}|^2),\\
F_{11}=(\mu_1+\nu_1)(|C_{01}|^2+|C_{10}|^2+2|C_{11}|^2)
+(\mu_2+\nu_2)(2|C_{00}|^2+|C_{01}|^2+|C_{10}|^2).
\end{eqnarray}
 We have set $\mu_1^a=\mu_1^b=\mu_1$,  $\mu_2^a=\mu_2^b=\mu_2$. 

In principle, (\ref{mat}) is a  basis for  two-bit operations.
For the sake of quantum computation, 
there are some   questions worthy of   investigations,
for example, whether  (\ref{mat}) can be used
to realize universal two-bit gates, 
whether there is universality for  nonlinear gates, 
how to construct  algorithms for  NP-complete and \#P problems 
based directly on (\ref{qun}) and  (\ref{mat}), 
how to realize  linear and simpler  two-bit operations  for 
the  Bose-Einstein condensates,  whether swapping operation~\cite{loss}
can be constructed,    etc.

\section*{6. Spinor  condensates}

\subsection*{6.1. Spinor condensate wavefunction}

Up to now, the internal state is irrelevant. In this
section,  in parallel to the above discussions on a condensate in a 
double-well potential,
we  discuss the Josephson-coupled internal states of a condensate. 
In this case, the  field operator
and  the condensate wavefunction
are    spinors.  We use a  
two-component condensate in an atom trap~\cite{williams} as the prototype
for  discussions.

Suppose that  the two internal states  are $|0\rangle$ and $|1\rangle$. Then
together with the motional degree of freedom, 
the single particle basis state can be written as 
$\phi_{\alpha,n}(\bbox{r})|n\rangle$, ($n=0,1$).
The field operator can be defined as 
\begin{eqnarray}
\hat{\psi}(\bbox{r},t)&=&
\sum_{\alpha}\sum_{n=0,1}\phi_{\alpha,n}(\bbox{r})|n\rangle
\hat{a}_{\alpha,n}(t)
\nonumber \\
&=& \hat{\psi}_0(\bbox{r},t)|0\rangle+\hat{\psi}_1(\bbox{r},t)|1\rangle, 
\nonumber \\ &\doteq& 
  \left( \begin{array}{c}
 \hat{\psi}_0(\bbox{r},t)\\
 \hat{\psi}_1(\bbox{r},t)
\end{array} \right), \label{fss}
\end{eqnarray}
where
$\hat{\psi}_n(\bbox{r},t)=
\sum_{\alpha}\phi_{\alpha,n}(\bbox{r})\hat{a}_{\alpha}(t)$.
Imposing  
SGSB  on  (\ref{fss}) leads to the spinor condensate wavefunction 
\begin{equation}
\Phi(\bbox{r},t) 
= \Phi_0(\bbox{r},t)|0\rangle+\Phi_1(\bbox{r},t)|1\rangle
\doteq 
 \left( \begin{array}{c}
 \Phi_0(\bbox{r},t)\\
 \Phi_1(\bbox{r},t)
\end{array} \right), \label{ssc}
\end{equation}
with $\Phi(\bbox{r},t)=\langle \hat{\psi}(\bbox{r},t) \rangle$, 
$\Phi_n(\bbox{r},t)= \langle \hat{\psi}_n(\bbox{r},t)\rangle$.  
  (\ref{ssc}) can be written as
\begin{equation}
\phi(\bbox{r},t) 
= \phi_0(\bbox{r},t)|0\rangle+\phi_1(\bbox{r},t)|1\rangle, \label{ssc2}
\end{equation}
with $\phi(\bbox{r},t)=\Phi(\bbox{r},t)/\sqrt{N}$,  
 $\phi_n(\bbox{r},t)=\Phi_n(\bbox{r},t)/\sqrt{N}$. We have used notations
 similar to those for the double-well condensate. But note the differences 
in meaning. 

The justification in terms of ODLRO can also be made.
The one-particle reduced density matrix is 
\begin{equation}
\langle \bbox{r}'|\hat{\rho}_1| \bbox{r}\rangle =  
|0\rangle\langle 0|
\langle\bbox{r}',0|\hat{\rho}_1|\bbox{r},0\rangle
+|0\rangle\langle 1|
\langle\bbox{r}',0|\hat{\rho}_1|\bbox{r},1\rangle
+|1\rangle\langle 0|
\langle\bbox{r}',1|\hat{\rho}_1|\bbox{r},0\rangle
+|1\rangle\langle 1|
\langle\bbox{r}',1|\hat{\rho}_1|\bbox{r},1\rangle,
\end{equation}
where $\langle\bbox{r}',n'|\hat{\rho}_1|\bbox{r},n\rangle
= \langle \hat{\psi}_{n'}(\bbox{r}',t)\hat{\psi}^{\dagger}_{n}(\bbox{r},t)
\rangle$. The existence of ODLRO implies
$\langle \bbox{r}',n'|\rho_1|\bbox{r},n\rangle
\rightarrow  \Phi_{n'}(\bbox{r}',t)\Phi_{n}^*(\bbox{r},t)$, and thus
\begin{equation}
\langle \bbox{r}'| \hat{\rho}_1| \bbox{r}\rangle \rightarrow 
(\Phi_0(\bbox{r}',t)|0\rangle+\Phi_1(\bbox{r}',t)|1\rangle)
(\Phi_0(\bbox{r},t)|0\rangle+\Phi_1(\bbox{r},t)|1\rangle)^*, 
\end{equation}
which indicates the existence of 
the spinor  condensate wavefunction $\Phi(\bbox{r})$,
as in  (\ref{ssc}).

\subsection*{6.2. Four-component 
spinor condensate wavefunction of two coupled  condensates }

For two coupled two-component   Bose systems $a$ and $b$, 
\begin{eqnarray}
\hat{\psi}(\bbox{r}^a,t)\hat{\psi}(\bbox{r}^b,t)
&=& \sum_{n^a}\sum_{n^b}\sum_{\alpha^a}\sum_{\alpha^b}
\phi_{\alpha^a,n^a}(\bbox{r}^a)\phi_{\alpha^b,n^b}(\bbox{r}^b)
|n^a\rangle|n^b\rangle \hat{a}_{\alpha^a,n^a}(t)\hat{a}_{\alpha^b,n^b}(t)
\nonumber 
\\  &=& \sum_{n^a}\sum_{n^b} 
\hat{\psi}_{n^a}(\bbox{r}^a,t)\hat{\psi}_{n^b}(\bbox{r}^b,t)
|n^a\rangle|n^b\rangle, \nonumber\\
&\doteq &
\left( \begin{array}{c}
 \hat{\psi}_{0}(\bbox{r}^a,t)\hat{\psi}_{0}(\bbox{r}^b,t)\\
 \hat{\psi}_{0}(\bbox{r}^a,t)\hat{\psi}_{1}(\bbox{r}^b,t)\\
 \hat{\psi}_{1}(\bbox{r}^a,t)\hat{\psi}_{0}(\bbox{r}^b,t)\\
 \hat{\psi}_{1}(\bbox{r}^a,t)\hat{\psi}_{1}(\bbox{r}^b,t)
\end{array} \right).
\label{xs}
\end{eqnarray}
Imposing SGSB on   (\ref{xs}), one obtains 
\begin{equation}
\Phi(\bbox{r}^a,\bbox{r}^b,t)
=\sum_{n^a=0,1}\sum_{n^b=0,1}
\Phi_{n^a,n^b}(\bbox{r^a},\bbox{r^b},t)|n^a\rangle|n^b\rangle
\doteq \left( \begin{array}{c}
 \Phi_{00}(\bbox{r}^a,\bbox{r}^b,t)\\
 \Phi_{01}(\bbox{r}^a,\bbox{r}^b,t)\\ 
\Phi_{10}(\bbox{r}^a,\bbox{r}^b,t)\\
\Phi_{11}(\bbox{r}^a,\bbox{r}^b,t)
 \end{array} \right),
\label{psi1s}
\end{equation}
where 
$\Phi(\bbox{r}^a,\bbox{r}^b,t)\equiv\langle \hat{\psi}(\bbox{r}^a,\bbox{r}^b,t)
\rangle$, 
$\Phi_{n^a,n^b}(\bbox{r^a},\bbox{r^b},t)\equiv 
\langle \hat{\psi}_{n^a,n^b}(\bbox{r}^a,\bbox{r}^b,t)\rangle$. 
Therefore  for two coupled two-component  condensates,
the total   condensate wavefunction is a  four-component spinor. 

The reasoning can also be cast in the form of
 ODLRO.   
The one-particle-pair reduced density matrix is  
\begin{eqnarray}
\langle {\bbox{r}^a}',{\bbox{r}^b}'|\hat{\rho}_1|\bbox{r}^a,\bbox{r}^b\rangle 
&\equiv &  \langle \hat{\psi}({\bbox{r}^a}',t)\hat{\psi}({\bbox{r}^b}',t) 
\hat{\psi}^{\dagger}(\bbox{r}^b,t)\hat{\psi}^{\dagger}(\bbox{r}^a,t)
 \rangle \nonumber  \\
&=& \sum_{{n^a}',{n^b}',n^a,n^b} 
\langle {\bbox{r}^a}',{n^a}',{\bbox{r}^b}',{n^b}'|\hat{\rho}_1|
\bbox{r}^a,n^a,\bbox{r}^b,n^b \rangle
|{n^a}'\rangle| {n^b}'\rangle\langle n^a|\langle n^b|, \label{o1s}
\end{eqnarray}
where 
\begin{equation}
\langle {\bbox{r}^a}',{n^a}',{\bbox{r}^b}',{n^b}'|\hat{\rho}_1|
\bbox{r}^a,n^a,\bbox{r}^b,n^b\rangle
=\langle \hat{\psi}_{{n^a}'}({\bbox{r}^a}',t)
\hat{\psi}_{{n^b}'}({\bbox{r}^b}',t)
\hat{\psi}_{n^b}^{\dagger}(\bbox{r}^b,t)
\hat{\psi}_{n^a}^{\dagger}(\bbox{r}^a,t) 
\rangle.  \label{odls}
\end{equation}
With ODLRO,
\begin{equation}
\langle \hat{\psi}_{{n^a}'}({\bbox{r}^a}',t)
\hat{\psi}_{{n^b}'}({\bbox{r}^b}',t)
\hat{\psi}_{n^b}^{\dagger}(\bbox{r}^b,t)
\hat{\psi}_{n^a}^{\dagger}(\bbox{r}^a,t) 
\rangle
\rightarrow
\Phi_{{n^a}',{n^b}'}({\bbox{r}^a}',{\bbox{r}^b}',t)
\Phi_{n^a,n^b}^*(\bbox{r}^a,\bbox{r}^b,t).
\end{equation}
Consequently,  (\ref{o1s}) approaches
$$[\sum_{{n^a}'=0,1}\sum_{{n^b}'=0,1}
\Phi_{{n^a}',{n^b}'}({\bbox{r}^a}',{\bbox{r}^b}',t)
|{n^a}'\rangle|{n^b}'\rangle]
[\sum_{n^a=0,1}\sum_{n^b=0,1}\langle n^a|\langle n^b|
\Phi_{n^a,n^b}^*(\bbox{r}^a,\bbox{r}^b,t)],
$$
implying  the existence of the
joint    condensate wavefunction as a four component spinor,
 as given in (\ref{psi1s}).  

We may write (\ref{psi1s}) as
\begin{equation}
\phi(\bbox{r}^a,\bbox{r}^b,t)
=\sum_{n^a=0,1}\sum_{n^b=0,1}
\phi_{n^a,n^b}(\bbox{r^a},\bbox{r^b},t)|n^a\rangle|n^b\rangle,
\label{psi3s}
\end{equation}
with $\phi(\bbox{r}^a,\bbox{r}^b,t)=\Phi(\bbox{r}^a,\bbox{r}^b,t)/N$,
$\phi_{n^a,n^b}(\bbox{r^a},\bbox{r^b},t)
=\Phi_{n^a,n^b}(\bbox{r^a},\bbox{r^b},t)/N$.

The many-body states are  similar to the cases in Sec. 3, only with
each single particle wavefunction $\phi_{\alpha,n}(\bbox{r})$ replaced as
the combination of motional state and the internal state 
 $\phi_{\alpha,n}(\bbox{r})|n\rangle$.

\subsection*{6.3. Hamiltonians and equations of motion}

When there is 
no coupling between the internal state and the motional state, 
$\phi_{\alpha,n}(\bbox{r})$ in (\ref{fss}) is independent of $n$.
 $\phi_{n}(\bbox{r},t)$ in (\ref{ssc2}) 
can be written as $c_n(t)\phi(\bbox{r},t)$,
where $\phi(\bbox{r},t)$ is independent of $n$, while
$c_n(t)$ is independent of $\bbox{r}$.
With an  electromagnetic field, 
the Hamiltonian for the  internal state is 
\begin{equation}
\hat{H}_{in}=\frac{\omega}{2}\hat{\sigma}_x+\frac{\delta}{2}\hat{\sigma}_z,
\label{one2}
\end{equation}
with $\hat{\sigma}_z|n\rangle=(2n-1)|n\rangle$.
 $\omega$ is the Rabi frequency,
while $\delta$ is the detuning. 
  
With the  coupling between the motional and
internal states, the total  Hamiltonian for a two-component 
 condensate  is 
\begin{equation}
\hat{H} = \sum_{i=1}^{N}[h(\bbox{r}_i)\otimes\hat{1}_{in}
+\hat{1}\otimes \hat{H}_{in}(i)] 
+\sum_{i<j} U(\bbox{r}_i-\bbox{r}_j),
\end{equation}
where, with the number of particles in each internal state
conserved, the potential $U$ is 
\begin{equation}
 U  \doteq
 \left( \begin{array}{cccc}
 U_{00} & 0 & 0 & 0 \\
 0     & \frac{1}{2}U_{01} & \frac{1}{2}U_{01} & 0 \\
 0     & \frac{1}{2}U_{01} & \frac{1}{2}U_{01} & 0 \\
0 & 0 & 0 & U_{11} 
 \end{array} \right),
\end{equation}
 in the basis  $|00\rangle$,  $|01\rangle$, $|10\rangle$, $|11\rangle$.
The field theoretical Hamiltonian  is 
\begin{eqnarray}
{\cal H}=\int d\bbox{r}\hat{\psi}^{\dagger}(\bbox{r})[\hat{h}(\bbox{r})
\otimes\hat{1}_{in}]\hat{\psi}(\bbox{r})+
\int d\bbox{r}\hat{\psi}^{\dagger}(\bbox{r})[\hat{1}\otimes \hat{H}_{in}]
\hat{\psi}(\bbox{r})\nonumber\\
+\frac{1}{2}\int \int d\bbox{r}_1d\bbox{r}_2 \hat{\psi}^{\dagger}(\bbox{r}_1)
\hat{\psi}^{\dagger}(\bbox{r}_2)U(\bbox{r}_1-\bbox{r}_2)
\hat{\psi}(\bbox{r}_2) \hat{\psi}(\bbox{r}_1), \label{h1t}
\end{eqnarray}
where the field operator $\hat{\psi}(\bbox{r})$ is as defined in (\ref{fss}).
The equation of motion of $\hat{\psi}(\bbox{r},t)$ is
\begin{equation}
i\hbar\frac{\partial\hat{\psi}(\bbox{r},t)}{\partial t}=
[(-\frac{\hbar^2\bigtriangledown^2_{\bbox{r}}}{2m}+V(\bbox{r}))
\otimes\hat{1}_{in}+\hat{1}\otimes \hat{H}_{in}
+\int \hat{\psi}^{\dagger}(\bbox{r}',t)
U(\bbox{r}-\bbox{r}') \hat{\psi}(\bbox{r}',t)
d\bbox{r}']\hat{\psi}(\bbox{r},t),
 \label{nsss}
\end{equation}
where a $2\times 2$ unit matrix is omitted in front of
$\hat{\psi}^{\dagger}(\bbox{r}',t)$ (similar is the following).  
The vertical position
of the  trapping potential   depends on the
internal state~\cite{williams}, 
\begin{equation}
V(\bbox{r}) \doteq diag[V_0(\bbox{r}),V_1(\bbox{r})],
\end{equation}
where  $V_n(\bbox{r})=\omega_{\rho}\rho^2/2
+\omega_{z}(z-z_{n})^2/2$. Therefore  we have 
\begin{equation}
i\hbar\frac{\partial}{\partial t}\left( \begin{array}{l}
 \hat{\psi}_0(\bbox{r},t)\\
 \hat{\psi}_1(\bbox{r},t) \end{array} \right)
= \left( \begin{array}{cc}
H_{00}+\delta/2   & \omega/2 \\
\omega/2 & 
 H_{11}-\delta/2 
\end{array} \right)
\left( \begin{array}{l}
 \hat{\psi}_0(\bbox{r},t)\\
 \hat{\psi}_1(\bbox{r},t) \end{array} \right).
 \label{ssss}
\end{equation}
where 
\begin{equation}
H_{nn}=
-\frac{\hbar^2\bigtriangledown^2_{\bbox{r}}}{2m}
+V_n(\bbox{r})+\sum_{n'=0,1}
\int |\hat{\psi}_{n'}(\bbox{r}',t)|^2 U_{nn'}(\bbox{r}-\bbox{r}')
d\bbox{r}',
\end{equation}
where $U_{10} \equiv U_{01}$.
Consequently, through SGSB or ODLRO,
 one obtains the  coupled Gross-Pitaevskii Equations for 
the spinor condensate wavefunction 
$\phi_n=\langle\hat{\psi}_n(\bbox{r})\rangle/\sqrt{N}$, 
\begin{equation}
i\hbar\frac{\partial}{\partial t}\left( \begin{array}{l}
 \phi_0(\bbox{r},t)\\
 \phi_1(\bbox{r},t) \end{array} \right)
= \left( \begin{array}{cc}
\overline{H}_{00}+\delta/2  & \omega/2 \\
\omega/2 & \overline{H}_{11} -\delta/2 
\end{array} \right)
\left( \begin{array}{l}
 \phi_0(\bbox{r},t)\\
 \phi_1(\bbox{r},t) \end{array} \right).
 \label{nss2}
\end{equation}
\begin{equation}
\overline{H}_{nn} = 
-\frac{\hbar^2\bigtriangledown^2_{\bbox{r}}}{2m}
+V_n(\bbox{r})+N\sum_{n'=0,1}\int 
|\phi_{n'}(\bbox{r}',t)|^2 U_{nn'}(\bbox{r}-\bbox{r}') d\bbox{r}'. \label{oco}
\end{equation}
If the interaction is s-wave interaction,  $U_{nn'}(\bbox{r}-\bbox{r}')
=g_{nn'}\delta (\bbox{r}-\bbox{r}')$. 

For  two coupled Bose systems $a$ and $b$,
the general  Hamiltonian is similar to Eq.~(\ref{many}), with 
each term now changed to  a matrix in the internal state  space, in a way
similar to the above  case of a single Bose system.
In terms of the  field operators $\hat{\psi}(\bbox{r}^a,t)$ and
$\hat{\psi}(\bbox{r}^b,t)$, formally
the field theoretical  Hamiltonian can   be  expressed  as  
(\ref{mn}), while the equation of motion is  in the form of 
(\ref{eom3}). 
However,  each potential or  interaction energy operator 
depends in the  appropriate way on the internal states.
In the matrix representation, 
\begin{equation}
V(\bbox{r}^i) \doteq diag[V_0(\bbox{r}^i),V_1(\bbox{r}^i)],
\end{equation}
in the basis $|0^i\rangle$, $|1^i\rangle$, ($i=a,b$), 
while it is assumed that 
\begin{equation}
 W(\bbox{r}^i-{\bbox{r}^j}') \doteq
diag[W_{00}(\bbox{r}^i-{\bbox{r}^j}'),W_{01}(\bbox{r}^i-{\bbox{r}^j}'),
 W_{10}(\bbox{r}^i-{\bbox{r}^j}'),W_{11}(\bbox{r}^i-{\bbox{r}^j}')]
\end{equation}
in the basis of $|0^i0^j\rangle$,  $|0^i1^j\rangle$, $|1^i0^j\rangle$,
 $|1^i1^j\rangle$, ($i,j=a,b$, $i\neq j$).  The long-range interaction
 within  a same condensate can be neglected, compared with $U$.  
Then it is easy to write Eqs.~(\ref{h0}) to  (\ref{eom3})
in the matrix form. The following is the equation of motion for
 $\hat{\psi}_{n^a}(\bbox{r}^a,t)\hat{\psi}_{n^b}(\bbox{r}^b,t) $.
\begin{equation}
\begin{array}{rcl}
i\hbar\frac{\partial[\hat{\psi}_{n^a}(\bbox{r}^a,t)
\hat{\psi}_{n^b}(\bbox{r}^b,t)]}{\partial t} &= & 
[-\frac{\hbar^2}{2m}\bigtriangledown_{\bbox{r}^a}^2+V_{n^a}(\bbox{r}^a)
-\frac{\hbar^2}{2m}\bigtriangledown_{\bbox{r}^b}^2+V_{n^b}(\bbox{r}^b)
\\
&& +\int d{\bbox{r}^a}'
\hat{\psi}^{\dagger}_{{n^a}'}({\bbox{r}^a}',t)
U_{n^a{n^a}'}(\bbox{r}^a-{\bbox{r}^a}')\hat{\psi}_{{n^a}'}({\bbox{r}^a}',t)
 \\
&&+\int d{\bbox{r}^b}'
\hat{\psi}^{\dagger}_{{n^b}'}({\bbox{r}^b}',t)
U_{n^b{n^b}'}(\bbox{r}^b-{\bbox{r}^b}')\hat{\psi}_{{n^b}'}({\bbox{r}^b}',t)
  \\
&&+\int d{\bbox{r}^a}'
\hat{\psi}^{\dagger}_{{n^a}'}({\bbox{r}^a}',t)
W_{{n^a}'n^b}(\bbox{r}^b-{\bbox{r}^a}')\hat{\psi}_{{n^a}'}({\bbox{r}^a}',t)
  \\
&&+\int d{\bbox{r}^b}'
\hat{\psi}^{\dagger}_{{n^b}'}({\bbox{r}^b}',t)
W_{n^a{n^b}'}(\bbox{r}^a-{\bbox{r}^b}')\hat{\psi}_{{n^b}'}({\bbox{r}^b}',t)]
\hat{\psi}_{n^a}(\bbox{r}^a,t)\hat{\psi}_{n^b}(\bbox{r}^b,t), 
\end{array}
\label{eoms}
\end{equation}
from which one obtains the equation of motion for the 
 condensate wavefunction component 
$\phi_{n^an^b}(\bbox{r}^a,\bbox{r}^b)$,
\begin{equation}
\begin{array}{rl}
i\hbar\frac{\partial \phi_{n^an^b}(\bbox{r}^a,\bbox{r}^b,t)}{\partial t} =
&[-\frac{\hbar^2}{2m}\bigtriangledown_{\bbox{r}^a}^2+V_{n^a}(\bbox{r}^a)
-\frac{\hbar^2}{2m}\bigtriangledown_{\bbox{r}^b}^2+V_{n^b}(\bbox{r}^b)
\\
&+N\int\int d{\bbox{r}^a}'d{\bbox{r}^b}'
(|\phi_{{n^a}'0}({\bbox{r}^a}',{\bbox{r}^b}',t)|^2 +
|\phi_{{n^a}'1}({\bbox{r}^a}',{\bbox{r}^b}',t)|^2) 
U_{n^a{n^a}'}(\bbox{r}^a-{\bbox{r}^a}')
\\
&+N\int\int d{\bbox{r}^a}'d{\bbox{r}^b}'
(|\phi_{0{n^b}'}({\bbox{r}^a}',{\bbox{r}^b}',t)|^2+
|\phi_{1{n^b}'}({\bbox{r}^a}',{\bbox{r}^b}',t)|^2)
 U(\bbox{r}^b-{\bbox{r}^b}')  \\
&+N
\int\int d{\bbox{r}^a}'d{\bbox{r}^b}'
(|\phi_{{n^a}'0}({\bbox{r}^a}',{\bbox{r}^b}',t)|^2 +
|\phi_{{n^a}'1}({\bbox{r}^a}',{\bbox{r}^b}',t)|^2)
W({\bbox{r}^a}'-\bbox{r}^b) \\
&+N
\int\int d{\bbox{r}^a}'d{\bbox{r}^b}'
(|\phi_{0{n^b}'}({\bbox{r}^a}',{\bbox{r}^b}',t)|^2+
|\phi_{1{n^b}'}({\bbox{r}^a}',{\bbox{r}^b}',t)|^2) 
W(\bbox{r}^a-{\bbox{r}^b}')] \phi_{n^an^b}(\bbox{r}^a,\bbox{r}^b,t). 
\end{array}
\label{couple2}
\end{equation}

In principle, the internal state may encode qubit. 
A simple  implementation of   one-bit linear operations can be made if
there is no coupling with the motional state.  Then applying an 
electromagnetic field can realize one-bit operations, based on 
Eq.~(\ref{one2}). There is no nonlinearity here,
since nonlinearity appears  only when the motional 
degree of freedom is involved.

Because for different bases internal states of a condensate in a trap, 
vertical positions are different, the inter-condensate interaction    
$W_{n^an^b}(\bbox{r}^a-\bbox{r}^b)$ depends on  $n^a$ and $n^b$. 
It can be arranged in such a way (Fig.~2) that 
$W_{00}(\bbox{r}^a-\bbox{r}^b)=W_{11}(\bbox{r}^a-\bbox{r}^b)$,
$W_{01}(\bbox{r}^a-\bbox{r}^b)=W_{01}(\bbox{r}^a-\bbox{r}^b)$.
Then conditional dynamics described above may be realized. 
However, it is difficult to use this to realize two-bit operations for the 
purpose of quantum computation. 
One might be tempted to realize conditional phases
coming from the motional degree of freedom, in a way similar  to
some proposals for individual atoms and ions~\cite{zoller1,zoller2}. 
However, the nonlinearity makes this scheme hard to be realized. 

\subsection*{7. Additional remarks}

[1] The effect of nonlinearity on the interference~\cite{niu}
needs more investigations, both from the perspective of 
physics and the perspective of   
quantum computation. In our scheme of entangling condensates, nonlinearity
appears in the two-bit evolution. One might  find an 
alternative way,  which does not introduce two-bit nonlinearity.
On the other hand, it is interesting
to study  computational issues in presence of nonlinearity.  
Even though two-bit nonlinearity is a disadvantage for quantum computing, the 
physics problem itself is still  interesting. 

[2] In the previously studied 
two-species condensate \cite{ho}, each  condensate  has
a single condensate wavefunction, so the issue of entanglement is out of
question. Moreover, if the two species are 
two different internal states of a same kind atom,  they can be
 Josephson-coupled.
A possible way of  realizing a  mixture with entanglement or
a four-component spinor condensate,  as discussed here,
is to  mix  two  different kinds of atoms, 
each with the  two internal states coupled through 
an electromagnetic field. 
The equation of motion of the joint  field can still be written as
Eq.~(\ref{eoms}), but the main interactions are all due to s-wave scattering, 
hence  $W_{n^an^b}=U_{n^an^b}$,
$U_{n^i{n^j}'}(\bbox{r}^i-\bbox{r}^j)=g_{n^i{n^j}'}
\delta(\bbox{r}^i-\bbox{r}^j)$, ($i,j=a,b$).

[3] An extreme  case  of the coupled many-particle systems  considered
here is a system of identical composite particles. Then the proposed 
situation of quantum computation 
becomes Bose condensation of identical quantum computers. 

[4] Superposition of  condensate wavefunctions might be  useful for quantum 
computation no matter whether the entanglement is realized and used.  

[5] If a swapping operation
 can be realized, then probably by using an
optical lattice trapping many condensates, an architect similar to the one 
in \cite{zoller2} may be constructed, where a  head qubit 
 mediates  operations between distant qubits. 

[6]  Many studies on Josephson junction between superconductors are 
based on quantizing a macroscopic Hamiltonian with macroscopic variables 
such as the  charge or particle number. This is  equivalent to the approach
of condensate wavefunction~\cite{leg}.
Therefore 
it is also a kind of  condensate qubit that  is used in the Josephson-junction
quantum  computation~\cite{shon,moij}. Nevertheless, in these
proposals, it is either the particle number or the phase   of the condensate 
wavefunction that encodes   the qubit.
In our proposal, it is the two branches
of the condensate wavefunction  that 
encode the bits.  Josephson-junction qubits, since they are
charged, were proposed  to be coupled via  an electric circuit.

\section*{8. Summary} 

Coherent properties of Bose-Einstein condensation 
are well described by  condensate wavefunction, which, in a mean field theory,
is the  single particle  state in which 
the condensation occurs. The many-body state is the product of  
this same state occupied by all the particles. 
In Josephson-like effect, the condensate wavefunction 
 is the superposition of two bases wavefunctions.
Here we go a step further to suggest that for two such condensates interacting
with each other, there is a joint condensate wavefunction
which is a superposition of the products of the bases wavefunctions of the
two condensate, hence
there can be entanglement.  The many-body  state is suggested to be
a product of copies of this joint condensate wavefunction, with
symmetrization.  

With  superposition and entanglement,
 condensate wavefunction, or macroscopic wavefunction, 
 may be used to implement 
quantum computation. That many identical particles occupy a same 
single particle state may lead to some intrinsic 
robustness and stability,
although there is an issue of phase diffusion of the condensate 
wavefunction. 
For better accuracy, one may try to realize
linear evolution of the  the condensate wavefunction. On the other hand, 
the nonlinearity due to  particle-particle interaction
may  turn out to be  a resource  of computational 
power, as indicated  in a  nonlinear quantum algorithm for NP-complete
problems~\cite{abram}.  
We have illustrated the ideas
by using Bose  condensation of trapped atoms, especially in
double-well potentials. 
We have  also discussed the existence of a 
four-component condensate wavefunction,
as a result of the coupling between two 
two-component condensates, each  of which has  two Josephson-coupled 
internal states. 

With these justifications, 
future researches may include  detailed calculations
under realistic physical conditions and the open issues mentioned above. 
It is also interesting to extend the consideration to other macroscopic
quantum coherent systems. It seems that our work in the meantime
is the first discussion 
on quantum entanglement between  ``second quantized''  many-particle systems
or quantum fields, with off-diagonal long-range order or
spontaneous symmetry breaking.  Through this work,
it is also seen   that new physics emergent  on a
new level of complexity may lead to new properties of computation. Indeed, 
``more is different''~\cite{and}.

\section*{Acknowledgement}
I am very grateful to Prof.~A.~J.~Leggett for enlightening and useful comments.
I also
thank   C. Kiefer,   S. Lloyd, G. Milburn and W. Zhang for 
 discussions.

\bigskip 

\section*{Appendix}

Here   we  derive   Gross-Pitaevskii equation  from  ODLRO.  
  
The one-particle reduced density matrix can be written 
as 
\begin{equation}
\langle \bbox{r}_1'|\hat{\rho}_1(t)|\bbox{r}_1\rangle
=\frac{1}{(N-1)!}\int\cdots\int d\bbox{r}_2 \cdots d\bbox{r}_N
\langle \bbox{r}_1',\bbox{r}_2,\cdots,\bbox{r}_N|\hat{\rho}(t)|
\bbox{r}_1,\bbox{r}_2,\cdots,\bbox{r}_N\rangle,
\end{equation}
from which we obtain 
\begin{equation}
i\hbar \frac{\partial}{\partial t}
\langle \bbox{r}_1'|\hat{\rho}_1(t)|\bbox{r}_1\rangle
=\frac{1}{(N-1)!}\int\cdots\int d\bbox{r}_2 \cdots d\bbox{r}_N
\langle \bbox{r}_1',\bbox{r}_2,\cdots,\bbox{r}_N|(\hat{H}\hat{\rho}-
\hat{\rho}\hat{H})|\bbox{r}_1,\bbox{r}_2,\cdots,\bbox{r}_N\rangle.
\end{equation}
One can obtain the equation of motion of $\hat{\rho}_1$ as 
\begin{eqnarray}
i\hbar \frac{\partial}{\partial t}
\langle \bbox{r}_1'|\hat{\rho}_1(t)|\bbox{r}_1\rangle
=\hat{h}(\bbox{r}_1')\langle \bbox{r}_1'|\hat{\rho}_1(t)|\bbox{r}_1\rangle
- \langle\bbox{r}_1'|\hat{\rho}_1(t)|\bbox{r}_1\rangle\hat{h}(\bbox{r}_1)
\nonumber \\
 +\int  d\bbox{r}_2[ V(\bbox{r}_1'-\bbox{r}_2)\langle \bbox{r}_1',\bbox{r}_2|
\rho_2| \bbox{r}_1,\bbox{r}_2\rangle-
\langle \bbox{r}_1',\bbox{r}_2|
\rho_2| \bbox{r}_1,\bbox{r}_2\rangle V(\bbox{r}_1-\bbox{r}_2)],
\end{eqnarray}
where $\langle \bbox{r}_1',\bbox{r}_2|\hat{\rho}_2|
\bbox{r}_1,\bbox{r}_2\rangle$ is the two-particle reduced density
matrix, to which 
the major contribution   comes from 
$\langle \bbox{r}_1'|\hat{\rho}_1|\bbox{r}_1\rangle
\langle\bbox{r}_2|\hat{\rho}_1|\bbox{r}_2\rangle$. Assuming
\begin{equation}
\langle \bbox{r}_1',\bbox{r}_2|\hat{\rho}_2|
\bbox{r}_1,\bbox{r}_2\rangle=\langle \bbox{r}_1'|\hat{\rho}_1|\bbox{r}_1\rangle
\langle\bbox{r}_2|\hat{\rho}_1|\bbox{r}_2\rangle, \label{id}
\end{equation}
and considering 
$ V(\bbox{r}_1-\bbox{r}_2)=g\delta(\bbox{r}_1-\bbox{r}_2)$,  and  ODLRO
as given in equation (\ref{od}), one obtains
\begin{eqnarray}
i\hbar\frac{\partial \phi(\bbox{r}_1',t)}{\partial t} \phi^*(\bbox{r}_1,t)
+ i\hbar\frac{\partial \phi^*(\bbox{r}_1,t)}{\partial t} \phi(\bbox{r}_1',t)
= \hat{h}(\bbox{r}_1')\phi(\bbox{r}_1',t)\phi^*(\bbox{r}_1,t)
- \hat{h}(\bbox{r}_1)\phi^*(\bbox{r}_1,t)\phi(\bbox{r}_1',t) \nonumber \\
+gN|\phi(\bbox{r}_1',t)|^2\phi(\bbox{r}_1',t) \phi^*(\bbox{r}_1,t)
-gN|\phi(\bbox{r}_1,t)|^2\phi^*(\bbox{r}_1,t) \phi(\bbox{r}_1',t), \label{gp0}
\end{eqnarray}
which leads to the Gross-Pitaevskii  equation. 

\newpage 

Figure captions:

Fig. 1. Two interacting Bose condensates.  Each condensate  
is trapped in a double-well potential, hence may represent a qubit. 
$|0\rangle$ is represented by the condensate wavefunction  at one well, while
$|1\rangle$ is represented by the condensate wavefunction 
 at the other well.  A two-bit operation between 
qubits $a$ and $b$ may be  based on  an  interaction 
 between atoms in  different condensates.
 
Fig2. Two interacting two-component Bose condensates. 
Each condensate has two Josephson-coupled internal states.
The trapping potentials  for   $|0\rangle$ and $|1\rangle$
are displaced between  each other, hence 
the interaction between   $a$ and $b$,
based on the long-range interaction between atoms in different
condensates,  depends  on the internal
  states of $a$ and $b$. Each internal 
state may be regarded as  a  bit. However,
due to coupling  with the motional states and the 
nonlinearity, it is difficult to use this situation to realize a
two-bit phase gate.  

\psfig{figure=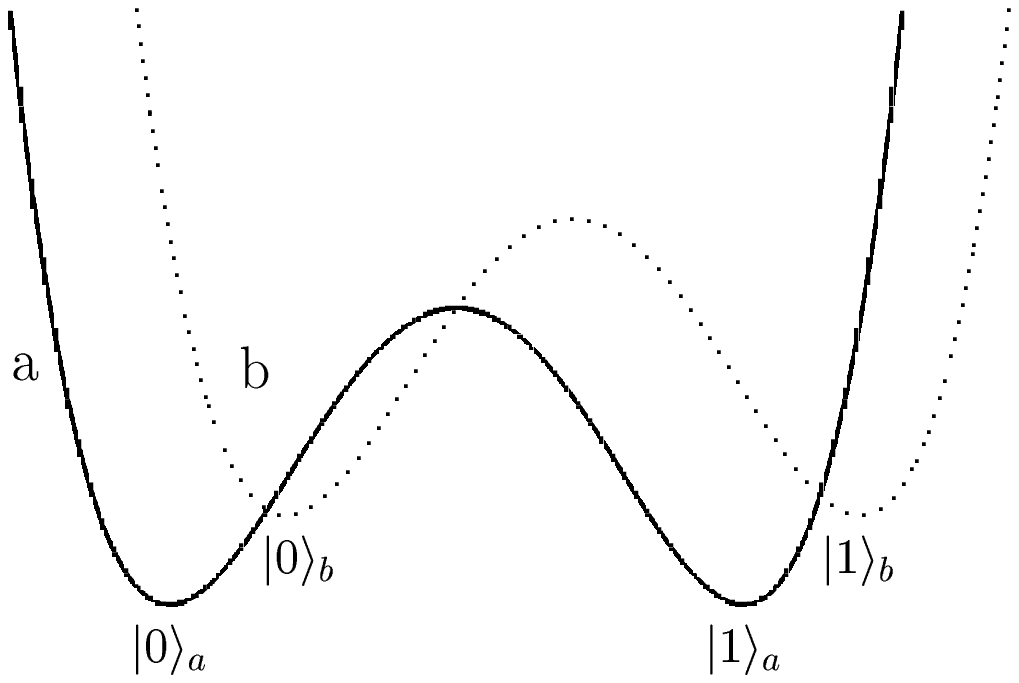}
\psfig{figure=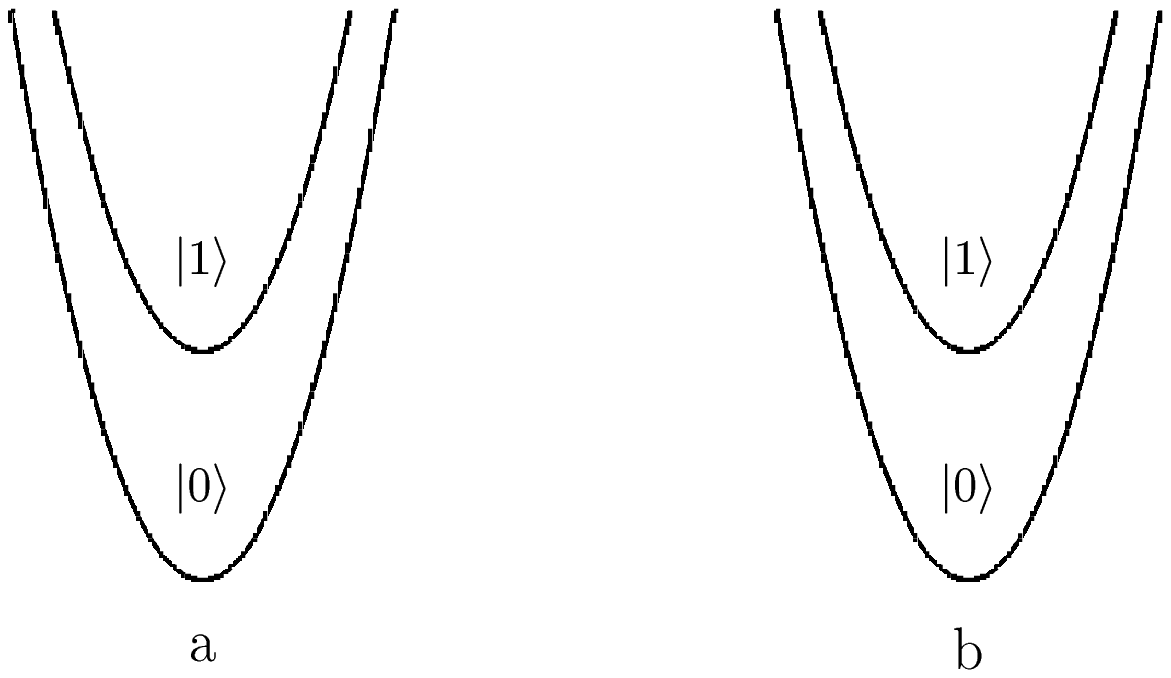}

\end{document}